\documentclass[]{article}

\usepackage{cite}

\usepackage[symbol]{footmisc}

\begin{document}

\begin{center}
{\Large{\bf {\sc Special Drawing Rights in a  New Decentralized Century\footnote{Paper proposal for the Georgetown University-International Monetary Fund Research and Policy Conference on Cryptoassets, 2018}}}}

\bigskip\medskip

Andreas Veneris\footnote{University of Toronto, Department of Electrical and Computer Engineering, and Department of Computer Science, 10 King's College Rd, Toronto, Ontario M5S 3G4, Canada, tel:~(416)~946-3062, veneris@eecg.toronto.edu} and Andreas Park\footnote{University of Toronto, Department  of Management and Rotman School of Management, University of Toronto, 105 St. George Street, Toronto, M5S 3E6, tel:~(416)~946-3015, andreas.park@rotman.utoronto.ca}

\end{center}


\bigskip
\noindent{\bf 1. Introduction}
\medskip

\noindent Unfulfilled expectations from macro-economic initiatives during the Great
Recession and the massive shift into globalization echo today  with   political upheaval, anti-establishment propaganda,  and looming trade/currency wars that threaten domestic and international value chains. 
Once stable entities like the EU now look  fragile and political instability in the US  presents unprecedented challenges to an  International Monetary System (IMS) that predominantly relies on the USD and EUR as reserve currencies \cite{Coats, IMFPolicy, Shelton, Xiaochuan}. 
In this environment, it is critical for an international organization mandated to ensure stability to plan and act ahead.
This paper argues that Decentralized Ledger-based Technology (DLT) is key for the  International Monetary Fund (IMF) to mitigate some of those risks, promote stability and safeguard world prosperity.
Over the last two  years, DLT has made headline news globally and created a worldwide excitement   not seen since the internet entered mainstream. 
The rapid adoption and ``open-to-all'' philosophy of DLT has already redefined global socioeconomics,
promises to shake up the world of  commerce/finance,  and challenges the workings of central governments/regulators. This paper examines DLT's core premises and proposes a two-step approach for the IMF to expand Special Drawing Rights (SDR)  into that sphere so as to become the originally envisioned numeraire and  reserve currency for cross-border transactions in this new decentralized century.


\vspace{0.2in}\noindent{\bf 2. The Global Socioeconomic Impact of Crypto-Economics}

\medskip \noindent The idea of  ``digital money'' is not new but early attempts failed because they could not solve  the double spending problem\cite{Digicash} --- until Bitcoin. 
 Building  cryptographic-economic principles  on a  peer-to-peer (P2P) network, Bitcoin:    {\em (i)} allows everyone to participate pseudo-anonymously, {\em (ii)} has a  market-driven  value, {\em (iii)} has low transaction fees that are determined by competition, {\em (iv)} allows  fast settlement with no  intermediaries, and {\em (v)} requires no  central authority with  jurisdiction over its operation.  
 As users  only need a smartphone to  transact in Bitcoin, it is not surprising that it has been  rapidly embraced in developing economies 
 that lack legacy telecommunication or proper financial infrastructure.

Ethereum generalized DLTs to materialize the vision of ``Decentralized Autonomous Organizations''~\cite{Szabo}. 
It is the world's first decentralized Turing Machine, a ``social operating system'' that  guarantees  trust in software  execution in terms of smart contracts through P2P consensus.  
Such contracts  enable commerce,  trading of financial securities, automated supply-chain management, enforcement/transfer of digital  rights, and transparent trade-offs between privacy and security~\cite{Moral}. Application sandboxes  are already found in Ukraine, which examines to use Ethereum to conduct an election~\cite{Ukraine}, Estonia, which develops a DLT-based e-residency to register out-of-country investments~\cite{Estonia} and Dubai's SmartCity, which awarded a blockchain contract to connect government and citizens~\cite{Dubai}. 
The latest cryptographic ``alt-coin'' DLTs such as Zcash, Monero, or ZK-Snarks (Ethereum)  ensure anonymity of network transactions, and protocols like Cardano and Algorand replace power wastage with  crypto-economic  consensus equilibria. 
Private permissioned blockchains such as Hyperledger and Corda use messaging to form endogenous DLT communities.   
 
Its cross-border nature complicates and confounds the regulation of DLTs~\cite{Blemus}, resulting  in 
 capital and technology flows to jurisdictions with favorable regulatory regimes. 
What remains remarkable  is that most DLT innovation was not  driven by firms that chase profits or governments that seek a national advantage ({\em i.e.,} for their military) but by freelance enthusiasts often coined as ``cypherpunks.'' 
Despite the ecosystem's significant volunteer nature, its ``distributed governance'' so far has proven resilient  and effective.

\vspace{0.2in} \noindent {\bf 3. SDR as Unit of Account and Secondary Markets}

\medskip\noindent
In 1969, the USD was pegged to gold and in light of ongoing looming changes in US money supply the international community felt it necessary to phase in   an alternative  international reserve asset.  
The IMF thus established the Special Drawing Rights (SDRs) and later, as part of  the  Second Amendment to its Articles, the Fund members agreed that SDRs will become the principal reserve asset in the IMS.

Much has changed in international finance since the 1970s as currencies are  no longer pegged to gold nor to one another. With exchange rates floating freely, the world economy has seen an unprecedented rise in trade and  economic prosperity. At the same time, international trade/finance remains anchored to the USD and, to a lesser extent, the Euro,  as primary reserve currencies. With this also came the Triffin dilemma, the international dependence on domestic US and EMU monetary policies. 
The recent decision of the US to start an international trade war to (presumably) address its domestic problems highlights once again the risk exposure of   world trade and international finance to internal
political shifts in  reserve currency countries. 

SDRs were originally intended (following the  Second Amendment) as the unit of account for international settlements.
An immediate advantage of the SDR's denomination as a basket of five currencies is that its {\em intrinsic} value depends less on a single country's political decisions allowing it to retain relative stability to any one of the currencies included in same basket. 
Common usage of SDRs would therefore reduce exposure to fluctuations in exchange rates and promote financial stability. For instance, since the basket expansion and US election in 2016, the brief historical data already show a convergence
of SDR-to-gold pricing in the 875-925 SDR/oz range, a much tighter zone  compared to  that of each currency in the SDR
basket  against the metal.  
Despite this attractive characteristic and the large expansion of allocations in 2009, today SDRs are still practically irrelevant in international trade chiefly due to geopolitical and institutional reasons.  
Migrating  SDR accounting to a permissioned DLT would promote wide adoption and help  SDRs to  finally fulfill their original mandate  of serving as the unit of account in international settlements.

Specifically, current and future accounting of SDRs should migrate to a  permissioned DLT with smart contracts~\cite{Ricardian} that: \textit{(i)} allow  automated and fully  transparent, passive allocation-and-redemption of SDRs based on balance of payment demands, \textit{(ii)} promote  the emergence of secondary private markets around SDRs (such as SDR-denominated bonds and derivatives), and \textit{(iii)} advocate SDR's role as the unit of account for international settlements. 
Recent projects by the Bank of Canada (Jasper) and the Monetary Authority of Singapore (Ubin) have set successful technical sandboxes around which such a DLT-based accounting system can be built. 
At the same time,   it is crucial that the new system is designed as forward-looking so as to ensure interoperability with existing permissionless DLTs (such as Ethereum) that can facilitate the creation of secondary private ``alt-SDR'' markets.

The benefits of a DLT-based SDR are multi-fold. 
Smart contracts are efficient mechanisms to execute  international transactions with an automated 2-step conversion through an in-and-out  process  ({\em e.g.,} Mexican Pesos to SDR to Danish Krone). 
By linking to an SDR platform, private businesses and the public can develop an automated, secure, and immutable ecosystem  to handle and improve existing multi-layer complex exchanges that  are cumbersome, slow, and excessively complex  for them  today with  multiple single-country reserve currencies. 
The key feature here is the availability of the multi-denominational SDR  and its stable intrinsic value, to promote  usage in the public and private sector with the goal of reducing uncertainty/costs/complexities in large- and small-scale cross-border transactions. 

As the acceptance of SDRs as a unit of exchange increases, there will be a natural increase of demand for SDR-related
financial instruments. 
As  recently witnessed within the Bitcoin and Ethereum ecosystems,  a plethora of innovative financial products are expected to emerge in  {\em private secondary markets}  such as ``alt-SDR'' denominated bonds, futures, or forwards.
This ``phased-in'' gradual development of a private ``alt-SDR''  security market  will further help to establish  SDRs as a formal numeraire in daily markets, without exerting abrupt pressure on the underlying basket of currencies, the
existing currency reserves, or the stability of the IMS.

Historically, ``hard'' currencies are often used to store value in countries that face political instability and large exchange rate/price fluctuations. In today's globalized economy, it is  imaginable that people who don't have access to a stable currency or worry about currency controls  resort to  using a ``world currency''  in denominating their daily commercial interactions. 
Ultimately, it is imaginable --and probably desirable--  that ``alt-SDRs''  become an easily accessible ``hard'' currency for eligible small businesses and individuals.

\vspace{0.2in}
\noindent {\bf 4. SDR as a Currency Reserve} 

\medskip \noindent 
History indicates that a move to pragmatically establish SDRs as a reserve  currency  may require exorbitant 
political will by the IMF's voting members. 
This paper argues that, in absence of acute economic turbulence, a thriving  private secondary market in SDR related instruments  will be the {\em catalyst}  to reach  agreement. 
The existence of such a market  may ultimately require amending the Articles of the IMF so as to allow: {\em (i)} the issuance of SDRs under currency board rules, {\em (ii)} the establishment of substitution SDR accounts of reserve assets  to provide liquidity, {\em (iii)}  the development of  further DLT-based financial instruments to ensure operation and transparency, and {\em (iv)} a gradual transition from the current basket of currencies  to a basket of common goods and commodities. 
The first two points have been debated  extensively in the prior literature, 
and we touched upon the third in the context of the preceding section. 
Although all four assertions are admittedly intertwined, in what follows we elaborate  on the last one.  

The purpose of SDRs is to facilitate trade-related international transactions and so instead of tying SDR to currencies that are subject to domestic monetary policies it is imaginable to tie its intrinsic value to a basket of commonly traded goods, commodities, and services that capture  real basic human needs. 
Such a  basket should  include metals, oil, and natural gas, all of which have shown to already  correlate well with present-day SDRs' value. 
It should also contain standardized agricultural products such as wheat,  corn, and soybeans. 
Finally, similar priority should be given in  pricing metrics such as a water cost index~\cite{IBM}, carbon emissions and
generation/use of renewable energy. 
Because human needs change only slowly over longer time horizons when compared to currency exchange rates, such anchoring is expected to create more stability yet respect global growth and promote prosperity.  
Blockchain technology can ensure that money supply is transparent; moreover,    all basket ingredients themselves would likely be tradable on financial markets as tokens, ensuring  arbitrage-free pricing, and existing central bank currencies would float freely against the SDR. 
The pegged basket of goods will become  the new ``gold'' standard, but without its shortcomings as it represents 
tangible modern human needs, and adheres to global agreed-upon constraints and requirements.

\vspace{0.2in}
\noindent {\bf 5. Conclusion}


\medskip\noindent The 
evolution of premissionless DLT platforms such  blockchain technology offers domestic and international monetary policy two choices: (1) disregard, regulate, and contain it so to maintain status quo, or (2) understand its intricacies and adapt existing practices~\cite{Lagarde}. The thesis of this paper is that macro-economic and political attitudes today place IMF at a central position to enhance SDR in a two-step process within a permissioned and permissionless DLT framework, shifting it closer to reserve currency status. 
Looking back at market charts of the music industry in the past 20 years indeed confirms that mishandling P2P innovation may backfire; only those who adapt lead the opportunity~\cite{IFPI}.

\bibliographystyle{unsrt}

\newpage

\begin{center}
{\bf    
 {\large {\sc Vita}}

(complete CVs in following pages)}
\end{center}

\vspace{0.35in}
{\bf Dr.\ Andreas Veneris} is a Connaught Scholar and Professor at the Department of Electrical and Computer Engineering, cross-appointed with the Department of Computer Science at the University of Toronto.  He is an alumni of the Japanese Society for the Promotion of Science hosted as visiting professor by the University of Tokyo (2010-11). In 2006-2016 he held a joint faculty position at Athens University of Economics and Business (Department of Informatics). He holds a PhD from the University of Illinois, Urbana-Champaign where he was also visiting faculty in 1998-99 before joining University of Toronto. His research is in formal methods for verification of smart contracts and systems, algorithms and crypto-economics, and ledger based technologies. He has published more than 130 papers in premier IEEE/ACM conferences/journals, he received a 10-year Best Paper Retrospective Award (IEEE/ACM Asian South Pacific Design Automation Conference, 2014), he holds multiple patents, and he was nominated for the Franklin Institute Bower Award and Price in Science in verification by Turing Award recipient Prof. Stephen Cook. Andreas has been involved with the vision and deployment of Ethereum along its founding team since 2013. He also founded the Blockchain Research Seminar Series at Fields Institute of Mathematics in 2017. In 2006, he led Vennsa Technologies in  Series A funding
to commercialize  research in formal methods serving Tier 1 semiconductor industry. In a previous life, he worked on the development of Mosaic (Netscape) and later he was member of the team that performed the first webcast ever ($37^{th}$ {\em Grammy Awards, March 1, 1995}), an event acknowledged in  the American Congress.  
 
\vspace{0.35in}

{\bf Dr.\ Andreas Park} has been an Associate Professor of finance at the University of Toronto since 2003, where he is a member of the Department of Management at the University of Toronto Mississauga,  the Rotman School of Management, and the Institute of Management and Innovation. Andreas holds a PhD in Economics from Cambridge University. He works on a number of theoretical and empirical research projects on the economic impact of technological transformations such as high frequency and dark trading in Canadian equity markets or blockchain technology on  design of markets for securities trading. He has received and has been an affiliate of a number of research grants from the ESRC in the UK, the SSHRC, the European Union, the Connaught Foundation, and the Global Risk Institute. His work has has been published in top journals in economics and finance such as Econometrica, the Journal of Finance, and the Journal of Financial Economics, and he is involved in several collaborative activities with Canadian securities regulators. 
Andreas has served as co-director of the Master of Financial Economics program at the University of Toronto; he is currently the Associate Chair of the Department of Management at UTM and also serves as the Research Director  at the Rotman School of Management's Financial Innovation Hub in Advanced Analytics.

\end{document}